\documentclass[prd,aps,tightenlines,superscriptaddress,nofootinbib,eqsecnum,amsfonts,amsmath,epsfig]{revtex4} 
\input{epsf} 
\usepackage{graphicx}

\newif\ifpdf
\ifx\pdfoutput\undefined
\pdffalse % we are not running PDFLaTeX
\else
\pdfoutput=1 % we are running PDFLaTeX
\pdftrue
\fi

\ifpdf
\usepackage{thumbpdf}
\usepackage{pdfsync}
\else
\usepackage{hyperref}
\fi

\def\be{\begin{eqnarray}}
\def\en{\end{eqnarray}}
\def\bea{\begin{eqnarray}}
\def\ena{\end{eqnarray}}
\def\di{\partial}
\def\cQ{{\cal Q}}

\begin{document}
\title{Mapping spacetimes with LISA: inspiral of a test-body in a `quasi-Kerr' field.}
  
\author{Kostas Glampedakis} 
\affiliation{School of Mathematics, University of Southampton, Southampton SO17 1BJ, UK}

\author{Stanislav Babak}
\affiliation{Max Planck Institute
(Albert Einstein Institute),
Am Muehlenberg 1, D-14476 Golm, Germany}
\date{\today}

\begin{abstract}

The future LISA detector will constitute the prime instrument for high-precision gravitational 
wave observations. Among other goals, LISA is expected to materialise a `spacetime-mapping' program, that is 
to provide information for the properties of spacetime in the vicinity of supermassive black holes which reside
in the majority of galactic nuclei. Such black holes can capture stellar-mass compact objects, which afterwards 
slowly inspiral under the emission of gravitational radiation. The small body's orbital motion and the associated 
waveform observed at infinity carry information about the spacetime metric of the massive black hole, and in principle 
it is possible to extract this information and experimentally identify (or not!) a Kerr black hole. 
In this paper we lay the foundations for a practical spacetime-mapping framework.
Our work is based on the assumption that the massive body is not necessarily a Kerr black hole, and
that the vacuum exterior spacetime is stationary-axisymmetric, described by a metric which deviates slightly from 
the known Kerr metric.  We first provide a simple recipe for building such a `quasi-Kerr' metric by adding to the Kerr metric the 
leading order deviation which appears in the value of the spacetime's quadrupole moment. We then study geodesic 
motion of a test-body in this metric, mainly focusing on equatorial orbits, but also providing equations describing generic orbits 
formulated by means of canonical perturbation theory techniques. We proceed by computing approximate `kludge' gravitational waveforms 
which we compare with their Kerr counterparts. We find that a modest deviation from the Kerr metric is sufficient for producing a significant 
mismatch between the waveforms, provided we fix the orbital parameters. This result suggests that an attempt to use Kerr waveform 
templates for studying extreme mass ratio inspirals around a non-Kerr object might result in serious loss of signal-to-noise 
ratio and total number of detected events. The waveform comparisons also unveil a `confusion' problem, that is the possibility of
matching a true non-Kerr waveform with a Kerr template of different orbital parameters.

\end{abstract}
\maketitle

%%%%%%%%%%%%%%%%%%%%%%%%%%%%%%%%%%%%%%%%%%%%%%%%%%%%%%%%%%%%%%%%%%

\section{Introduction}

During the past several years astronomical observations have provided almost indisputable evidence in favour of 
the existence of `dark' supermassive objects (with a mass spectrum $\sim 10^5 - 10^9 M_{\odot}$) in the majority 
of galactic cores, including our own Milky Way \cite{MBHs}. Conventional wisdom dictates that these objects 
should be Kerr black holes as described by General Relativity. This belief is (so far) rather based 
on our faith in General Relativity itself rather than on hard `experimental' evidence \footnote{Note however that the existence
of an event horizon seems to be required in order to explain the spectral properties of X-ray emission by Galactic black hole binaries and
AGNs, see for example Ref.~\cite{Xray}}. 
Other candidate massive objects have been proposed (such as soliton stars, boson stars, gravastars and P-stars \cite{others}) 
but these are treated with scepticism by the community as they involve `exotic' physics.  

The future LISA gravitational wave detector \cite{LISA} is expected to be able to give a definitive answer 
to whether these massive objects are Kerr black holes or not. It is commonly accepted \cite{kip} that LISA 
will have the potential of `mapping' the spacetime in these objects' close vicinity. This can be achieved by 
detection of gravitational radiation emitted during the inspiral of a stellar-mass compact body around the 
supermassive object. It is expected that LISA will be capable of detecting from several to about a thousand \cite{rates} 
of these extreme mass ratio inspirals (EMRIs), during a $3-5$ years mission. For the actual detection and subsequent extraction of 
the system's physical properties, LISA's data analysis will rely heavily on matched filtering. 
This method is based on cross-correlating the detector's noisy output with a pre-selected bank of waveform templates 
which should be accurate representations of the true signal. Among anticipated gravitational wave sources, EMRIs 
are thought of as one of the `cleanest': they can be accurately modelled as a binary system of a `test-body' orbiting a Kerr 
black hole, evolving under its own gravitational dynamics, without receiving any other significant 
`environmental' influence \footnote{This should be true for the majority of galactic supermassive black holes which 
are in a ``quiescent'' accretion state. For the small fraction of strongly accreting black holes (like the ones in AGNs) the presence
of an accretion disk may have non-negligible effects on a orbiting body (see for example \cite{chakrabarty})}. 

The basic tool for studying such a system is 
black hole perturbation theory: the small body of mass 
$\mu \sim 1-10 M_{\odot}$ perturbing the spacetime of the 
much more massive 
black hole (of mass $M \sim 10^6 $). For recent detailed reviews on EMRIs and further references on the subject we refer the 
reader to Refs.~\cite{kg_review}, \cite{chapter}.

In their great majority, EMRI-related studies take for granted the Kerr identity of the supermassive object. However, if we ever hope to 
materialise any spacetime-mapping program with LISA we should be prepared to dismiss (at least to some degree) this assumption. In fact,
any claim that LISA will be able to probe the Kerr metric would require the use of realistic waveform templates which contain 
--in some suitable parametric form-- deviations from the Kerr metric. The task of a LISA data analyst would be to quantify
these deviations, most likely pointing to a null result.

This idea was first put forward by Fintan Ryan in the late 90s. In a series of papers \cite{ryan}, Ryan 
demonstrated how LISA could in principle construct a map of the massive body's spacetime by means
of EMRIs observations. His method is based on writing the general stationary axisymmetric vacuum metric in terms 
of multipole moments as prescribed in Ref.~\cite{moments}. These can be mass moments $M_\ell$ and current moments $S_\ell$, both 
labelled by the angular integer eigenvalue $\ell \geq 0$. Given the moments, the metric takes the {\em symbolic} form 
\cite{ryan},\cite{moments},
\be
g_{ab} \sim \sum_{n=0}^{\infty} \left [ \frac{1}{r^{2n+1}}\, \left \{ M_{2n} {\cal P}^{1}_{n}(\theta) + 
(M_{\ell < 2n}, S_{\ell < 2n}) \right \},
\frac{1}{r^{2n+3}}\, \left \{ S_{2n+1} {\cal P}^{2}_{n}(\theta) +(M_{\ell < 2n+1}, S_{\ell < 2n+1})    
\right \} \right ] 
\label{mexpan}
\en
where ${\cal P}^{1,2}_{n}(\theta)$ are known angular functions and angular brackets stand for the contribution
of lower moments (if they exist for given $n$).    

These multipole moments will be clearly encoded in the geodesic equations of motion\footnote{Multipolar expansion of the 
gravitational potential is a powerful technique in Newtonian gravity for performing geodesic 
measurements: the motion of a satellite around the Earth provides information
on our planet's `bumpiness'.} for a test-body and consequently in the phase
and amplitude of the emitted waveform. LISA (as any other gravitational wave detector) will be capable of 
tracking this phase with high precession and extract valuable information concerning the spacetime's multipolar 
structure.  

The key feature that makes this method so attractive is that Kerr spacetime is special, in the sense that
all of its higher moments are 'locked' to the first two, the mass $M = M_0$ and spin $J=S_1$ \cite{moments},  
\be
M_\ell + i\,S_\ell = M(ia)^\ell
\label{nohair}
\en
where $a = J/M$ is the familiar Kerr spin parameter. 
This relation is unique to Kerr holes and is a mathematical expression of the famous `no-hair' 
theorem \cite{MTW}. In practise this means that extracting just the three lowest moments, $M,J$ and (say) 
the quadrupole moment $M_2$, is sufficient for verifying (or disproving!) that the central body is indeed a 
Kerr black hole. Measurement of additional higher moments could pinpoint the identity of the body 
\footnote{For example, a rotating massive Boson star \cite{ryan_boson} has three independent moments and its higher 
moments certainly do not obey relation (\ref{nohair}).}. We should emphasise that a possible non-Kerr multipolar
structure does not necessarily imply a non-Kerr identity for the massive object. A true Kerr hole with a substantial  
amount of material in its vicinity (say, in the form of an accretion disk) could effectively behave as a
non-Kerr object. This scenario is not included in our present discussion but certainly merits a future
detailed investigation.      

Ryan's multipole expansion scheme is certainly an elegant and powerful method for `mapping spacetimes', 
nevertheless there are some serious issues regarding its practicality. Computing gravitational 
waveforms for EMRIs requires (i) precise knowledge of a test-body's geodesic motion and (ii) a wave-emission
formalism that allows the calculation of waveforms and fluxes, once the orbital motion is prescribed.
For the case of a Kerr black hole, geodesic motion is well known and extensively studied \cite{chandrabook}, while wave
dynamics is studied with the help of the celebrated Teukolsky formalism \cite{teuk} which encapsulates the dynamics 
of gravitational perturbations in one single master equation. However, it is not always appreciated that the study
of both orbital and wave dynamics is greatly simplified due to the speciality of Kerr spacetime. In particular,
one is able to first decouple and then separate the dynamical equations thanks to the Petrov type-$D$ character 
of the Kerr spacetime and a suitable choice of coordinate frame \cite{stewart},\cite{carter}. 

Life is considerably more complicated if we consider a non-Kerr spacetime with arbitrary multipolar 
structure. To begin with, according to (\ref{mexpan}) an accurate representation of the metric 
in the strong field regime (the most relevant part of the inspiral for LISA) requires the inclusion of a 
large number of multipoles. This point can be made clearer if we notice that the metric (\ref{mexpan})
is an expansion in $1/r$ around the flat Minkowski spacetime. A more suitable expansion would be around
(say) the Schwarzschild spacetime. We will return to this point in future work.

Complications first appear at the level of geodesic motion. The spacetime's stationary-axisymmetric character guarantees 
the existence of two integrals of motion, the energy $E$ and angular momentum (along the symmetry axis) $L_z$. 
However, the third integral of motion (the famous Carter constant for the Kerr metric) will typically be lost, as the Hamilton-Jacobi 
equation is not fully separable (see \cite{MTW} for the solution of this equation in the Kerr case). Two integrals of motion 
are sufficient (together with initial conditions) for fully describing motion only for the special case of
equatorial orbits. For generic orbits (which is the case relevant for LISA) the insufficient number of integrals of motion
would force us to use the second-order geodesic equation of motion. This would result in a more complicated description
of geodesic motion (as compared to the Kerr case), but technical complication is not the sole problem here. Fewer integrals
of motion could easily translate into chaotic dynamical behaviour which in turn would have a great impact on 
LISA data analysis. 

The picture appears even more blurred when one attempts to study wave dynamics in a non-Kerr metric.  
Lack of decoupling for the perturbation equations inhibits the formulation of a `Teukolsky-like' equation -- 
the most versatile method in black hole perturbation theory is now lost. An alternative way of computing 
rigorous gravitational waveforms and fluxes would be the direct numerical time evolution of metric 
perturbations (after separation of the azimuthal $\phi$ coordinate) which come in a package of
ten coupled partial differential equations (combined with four gauge-condition equations) \cite{MTW}.   

Motivated (and partially discouraged!) by the above complications, we adopt a somewhat 
different (albeit less general) approach in the present work. Based on the belief that the massive objects in galactic nuclei 
are most likely Kerr black holes, we only address the question: is the spacetime around these objects described 
by the exact Kerr metric, or by a {\em slightly} different metric ? We do {\it not} attempt to identify the
source of this metric, in fact we are agnostic on this issue. Our less ambitious goal is to quantify the deviation 
from the Kerr metric. 

This point should be emphasised because none of the other candidate objects can be considered in any sense as being `almost'
a Kerr black hole. The idea of building a `quasi-black hole' spacetime has been recently advocated by
Collins \& Hughes \cite{hughes}. These authors constructed a `bumpy' Schwarzschild black hole by 
adding a certain amount of quadrupole moment (in the form of a given mass distribution outside
the black hole) and then studied equatorial orbits in the resulting spacetime. 

In our scheme we modify the quadrupole moment of the exact Kerr metric (this is the lowest moment where
Kerr can be distinguished from the spacetime of another axisymmetric/stationary body).
This procedure is performed in a natural way with the help of the well-known Hartle-Thorne exterior metric \cite{HT}. 
The resulting `quasi-Kerr' metric is customised for use in strong gravity situations
and has the nice feature of smoothly reducing to the familiar Kerr metric in Boyer-Lindquist 
(BL) coordinates. Being arbitrarily close to the BL Kerr metric (which allows full separability of the Hamilton-Jacoby 
and wave equations) allows us to formulate first-order equations of motion and integrals of motion by perturbing the 
well known Kerr equations. As yet, we have {\em not} explored whether the same trick would lead to a `quasi-Teukolsky' 
perturbation equation. For the purposes of the present investigation it is quite sufficient to use the so-called
hybrid approximation \cite{GHK},\cite{kludge_paper} and compute `kludge' waveforms.

The paper is structured as follows. In Section~\ref{sec:building} we provide the prescription for building
our quasi-Kerr metric. This is the paper's main result. Section~\ref{sec:geodesic} is devoted to the solution of the Hamilton-Jacobi 
equation in the quasi-Kerr metric. We first discuss the issue of separation of variables for this equation and then proceed
with the derivation of equations of motion based on standard canonical perturbation theory.
Section~\ref{sec:equat} is dedicated to the study of equatorial orbits and comparison of orbital frequencies and 
periastron advance for Kerr and quasi-Kerr orbits. In Section~\ref{sec:waveforms} we compute approximate waveforms from 
test-bodies in equatorial orbits of the quasi-Kerr metric. These are compared to their Kerr analogues by calculating 
their overlap function in the LISA sensitivity band. In the same Section we touch upon the problem of `confusion' between Kerr and 
quasi-Kerr waveforms. A summary and concluding discussion can be found in Section~\ref{sec:conclusions}. Appendices containing some 
technical details, including a review of the action-angle formalism in the Kerr spacetime can be found at the end of the paper. 
Throughout the paper we use geometrised units $G=c=1 $ and adopt a $\{-,+,+,+\}$ signature for the metric. We repeatedly use the 
labels `K' and `qK' for `Kerr' and `quasi-Kerr', respectively.

%%%%%%%%%%%%%%%%%%%%%%%%%%%%%%%%%%%%%%%%%%%%%%%%%%%%%%%%%%%%%%%%%%%%%%%%%%%%%%%%%%%%%

\section{Building a quasi-Kerr spacetime}
\label{sec:building}

As we mentioned in the Introduction, the central idea of our work is to replace the general 
multipolar expansion (\ref{mexpan}) of an axisymmetric-stationary vacuum spacetime with a less general
`quasi-Kerr' metric. In the language of multipole moments this means,
\bea
M_\ell &=& M_\ell^{\rm K} + \delta M_\ell, \quad \ell \geq 2
\\
S_\ell &=& S_\ell^{\rm K} + \delta S_\ell, \quad~~ \ell \geq 3
\ena
where $\delta M_\ell, \delta S_\ell $ are assumed small. With respect to the first two leading moments $M,J$  the spacetimes outside a 
Kerr hole and any other stationary axisymmetric rotating body are indistinguishable. Essentially, $M$ and $J$ can take the same 
value for both bodies (the spin is presumably limited, for a Kerr hole $a \leq M$). This degeneracy is broken as soon as we take 
into account the next most important moment, the mass quadrupole $M_2$. In our scheme we choose to take into account only the deviation 
in the quadrupole moment $ \cQ \equiv M_2 $ and {\it neglect} any deviation in all higher moments.  We introduce the dimensionless 
deviation parameter $\epsilon $ as \footnote{To convey an idea on the range of values that $\epsilon$ could take, we use the results
of Laarakkers \& Poisson \cite{PoisNS} for the quadrupole moment of rotating neutron stars. They show that the quadrupole 
moment could be approximated as $q \approx b j^2$. The factor $b$ is a function of the mass and equation of state, with numerical 
values varying from 2 to 12 (decreasing with mass and increasing with stiffness of the equation of state). It was 
emphasised in \cite{PoisNS} that this relationship holds even for fast rotations. The fact that $b$ is positive reflects oblateness 
of the mass distribution. Using this relationship we obtain: $\epsilon = (b-1)\,j^2$,
which tells us that for (stiff) rotating neutron star (with $M=1.0$) $\epsilon$ could be as large as $11\,j^2$.}

\be
\cQ = \cQ^{\rm K} - \epsilon\,M^3, \quad \mbox{where} \quad \cQ^{\rm K} = -\frac{J^2}{M}. 
\en
Our quasi-Kerr metric takes the form,   
\be
g_{\alpha\beta} = g_{\alpha\beta}^{\rm K} + \epsilon h_{\alpha\beta}  + 
{\cal O}(\delta M_{\ell \geq 4}, \delta S_{\ell \geq 3})
\label{qKerr}
\en
where $g_{\alpha\beta}^{\rm K} $ is the {\it exact} Kerr metric. Our goal is to determine the metric 
functions $h_{\alpha\beta}(r,\theta)$.  

This can be achieved by using the Hartle-Thorne (H-T) metric \cite{HT} which describes the spacetime outside {\it any} 
slowly rotating axisymmetric and stationary body. Indeed, this metric is fully accurate up to the quadrupole moment 
(which scales quadratically with the body's spin), and it includes as a special case the Kerr metric 
(at ${\cal O}(a^2)$ order). It is then possible to isolate the leading-order quadrupole moment deviation 
and deduce the $h_{\alpha\beta}$ piece in (\ref{qKerr}). Implicit in this prescription is the assumption that the massive body's exterior 
spacetime has nonzero higher moments induced by the body's rotation, in the same sense that a black hole has nonzero moments 
(with $\ell > 0  $) only when $ a \neq 0 $.  As a consequence, terms of order ${\cal O}(\epsilon a,\epsilon^2)$ are
beyond the desired accuracy and therefore neglected. Unlike the general multipole expansion (\ref{mexpan}), our prescription 
by construction offers no information (regarding possible deviations from the Kerr values) on moments higher 
than the quadrupole. On the other hand our metric is not an expansion in inverse powers of $r$ , instead is
written in a compact fully relativistic form suitable for strong-field conditions. 

We move on to the calculation of $h_{ab} $. For our purposes we shall only need the {\it exterior } H-T metric \cite{HT} 
which is expressed in terms of the dimensionless parameters,
\be
j \equiv \frac{J}{M^2}, \qquad q \equiv -\frac{\cQ}{M^3}
\en
Then at ${\cal O}(j^2)$ accuracy, the metric is given by (these expressions are taken from Ref.~\cite{abramowicz} after 
correcting a typographical sign error in the $F_1$ function), 
\bea
g_{tt}^{\rm HT} &=& - \left ( 1-2M/r \right ) [\, 1+j^2 F_1 + q F_2 ]
\nonumber \\
\nonumber \\
g_{rr}^{\rm HT} &=& \left ( 1 - 2M/r \right )^{-1} [\, 1+ j^2 G_1 -q F_2 ]
\nonumber \\
\nonumber \\
g_{\theta\theta}^{\rm HT} &=& r^2 [\, 1+ j^2 H_1 + q H_2 ]
\nonumber \\
\nonumber \\
g_{\phi\phi}^{\rm HT} &=& r^2 \sin^2\theta\, [\, 1 + j^2 H_1 + q H_2 ]
\nonumber \\
\nonumber \\
g_{t\phi}^{\rm HT} &=& -\frac{2M^2 j}{r} \sin^2 \theta
\label{HT1}
\ena
The functions $F_{1,2}(r,\theta),~H_{1,2}(r,\theta)$ appearing in these expressions are given in Appendix~\ref{app:HT_funcs}. 
Note that the metric (\ref{HT1}) is written in the coordinate frame $\{t,r,\theta,\phi\} $ originally used by Hartle \& Thorne. 
Without loss of generality, we can write
\be
q = j^2 + \epsilon
\label{deviation}
\en 
Clearly, the limit $\epsilon \to 0$ corresponds to the H-T representation of Kerr spacetime.

For reasons we discuss later in the paper, it is highly desirable for the quasi-Kerr metric to reduce 
to the familiar Kerr metric in Boyer-Lindquist (BL) coordinates in the limit $\epsilon \to 0 $. 
The transformation between BL and the original H-T coordinates was provided by Hartle \& Thorne \footnote{Note 
that both the original transformation equations by Hartle \& Thorne and 
their recent reproduction in Ref.~\cite{abramowicz} contain a sign error.}:
\bea
t_{\rm BL} &=& t, \qquad \phi_{\rm BL} = \phi
\\
\nonumber \\
r_{\rm BL} &=& r - \frac{j^2 M^2}{2r^3} [ (r+2M)(r-M) -\cos^2\theta (r-2M)
(r+3M) ] 
\nonumber \\
%\nonumber \\
&\equiv&  r - j^2 M^2 f_{\rm BL}(r,\theta)
\\
\nonumber \\
\theta_{\rm BL} &=& \theta -\frac{j^2 M^2}{2r^3} (r+2M) \sin\theta
\cos\theta \; \equiv \theta -j^2 M^2 g_{\rm BL}(r,\theta)
\label{HT2BL}
\ena
Written in this manner, this transformation is not really practical for someone wishing to obtain the metric in BL coordinates. 
Fortunately, inverting the above equations with respect to the H-T coordinates is a trivial task. According to (\ref{HT2BL})
$ r_{BL} = r + {\cal O}(j^2)$ and $ \theta_{BL} = \theta + {\cal O}(j^2)$; therefore we are free to use either set of coordinates 
in the functions $f_{BL}, g_{BL}$ since these appear in the ${\cal O}(j^2)$ terms. Thus, the
inverse transformation is simply,
\bea
r &=& r_{\rm BL} + j^2 M^2 f_{\rm BL}(r_{\rm BL},\theta_{\rm BL})
\nonumber \\
%\nonumber \\
\theta &=& \theta_{\rm BL} + j^2 M^2 g_{\rm BL}(r_{\rm BL},\theta_{\rm BL})
\label{BL2HT}
\ena   
Hereafter, we shall drop the subscript on the BL coordinates. As the two sets of coordinates differ at order ${\cal O}(j^2) $
the quadrupole-order pieces in (\ref{HT1}) are immune with respect to the transformation (\ref{BL2HT}). Only the `Schwarzschild' 
portion of the metric is affected by the coordinate transformation. For the contravariant metric components we obtain:
\bea
g^{tt}_{\rm HT} &=& g^{tt}_{\rm Ka^2} + \epsilon\, (1-2M/r)^{-1} [\, f_3(r)
+ f_4(r) \cos^2\theta ] 
\nonumber \\
\nonumber \\
g^{rr}_{\rm HT} &=& g^{rr}_{\rm Ka^2} + \epsilon\, (1-2M/r) [\, f_3(r) 
+ f_4(r) \cos^2\theta ] 
\nonumber \\
\nonumber \\
g^{\theta\theta}_{\rm HT} &=& g^{\theta\theta}_{\rm Ka^2} -\frac{\epsilon}{r^2}\,
[\, h_3(r) + h_4(r) \cos^2\theta ] 
\nonumber \\
\nonumber \\
g^{\phi\phi}_{\rm HT} &=& g^{\phi\phi}_{\rm Ka^2} -\frac{\epsilon}{r^2\sin^2\theta}\,
[\, h_3(r) + h_4(r) \cos^2\theta ] 
\nonumber \\
\nonumber \\
g^{t\phi}_{HT} &=& g^{t\phi}_{\rm Ka^2},
\label{HT_BL1}
\ena
where the functions $f_3, f_4, h_3, h_4$ are given in Appendix~\ref{app:HT_funcs}.
In eqns.~(\ref{HT_BL1}) we have denoted as $g^{\alpha\beta}_{\rm Ka^2}$ the ${\cal O}(a^2)$ Kerr 
metric (in BL coordinates) with $ a = jM $. From eqns.~(\ref{HT_BL1}) one can extract 
the desired deviation $h_{\alpha\beta}$.

Putting all the pieces together, the ansatz for building a quasi-Kerr metric in Boyer-Lindquist
coordinates is:
\be
g_{ab} = g_{ab}^{\rm K} + \epsilon\,h_{ab}
\label{qmetric}
\en        
where,
\bea
h^{tt} &=& (1-2M/r)^{-1} [\, (1-3\cos^2\theta)\,{\cal F}_1(r)] , \qquad 
h^{rr} = (1-2M/r) [\, (1-3\cos^2\theta)\,{\cal F}_1(r)]  
\\
\nonumber \\
h^{\theta\theta} &=&  -\frac{1}{r^2}\, [\, (1-3\cos^2\theta)\, {\cal F}_{2}(r)], \qquad
h^{\phi\phi} =  -\frac{1}{r^2\sin^2\theta}\,[\, (1-3\cos^2\theta)\, {\cal F}_{2}(r)] 
\\
\nonumber \\
h^{t\phi} &=& 0
\ena
The functions ${\cal F}_{1,2}(r) $ are given in Appendix~\ref{app:HT_funcs}.

Looking at the explicit form of $h_{\alpha\beta} $ it is evident that these functions are divergent at 
$r \to 2M $. This would correspond to the location of the event horizon at the accuracy ${\cal O}(j^2) $ of the
H-T metric. Hence, there is no real mystery behind this behaviour which is also present in the case of the
full Kerr metric. However, since in our scheme we do not associate the quasi-Kerr metric with a black hole 
we are forced to be agnostic regarding the nature of the `surface' of the massive body. Whether is an event
horizon or something else is an issue we simply do {\it not} address. 
This lack of information has no significant impact on the study of orbital motion of a test-body, as long as we do not 
consider orbits too close to $2M $ (assuming they exist).

The situation is not so clear when one tries to study wave dynamics in a quasi-Kerr metric and compute (say) fluxes.
To begin with, we should mention that a first approximation would be to completely neglect any gravitational flux impinging on 
the  `surface', assuming that is small compared to the fluxes at infinity. This is certainly true for the case of black holes 
\cite{kgdk},\cite{scott} and it is not unreasonable to expect the same in other scenarios. The main issue regards the computation of 
the fluxes at infinity, and their sensitivity to the absence of a concrete boundary condition at $ r \sim 2M $. One could try different 
boundary conditions, say, perfect reflection or free propagation and assess how robust the flux results are (see for example
Ref.~\cite{ryan}). Other aspects of wave dynamics in a quasi-Kerr field, such as the existence and properties of quasi-normal modes, 
are extremely sensitive to boundary conditions and cannot be studied within our framework.

%%%%%%%%%%%%%%%%%%%%%%%%%%%%%%%%%%%%%%%%%%%%%%%%%%%%%%%%%%%%%%%%%%%%%

\section{Geodesic motion in the quasi-Kerr metric}

\label{sec:geodesic}

\subsection{The Hamilton-Jacobi equation: issues of separability}
\label{sec:separability}

An elegant method for deriving equations of motion for a point-particle in a given
gravitational potential is the Hamilton-Jacobi (H-J) formalism (see Ref.~\cite{goldstein} 
for background material and Appendix~\ref{app:KerrHJ} of the present paper). 
The end-product is the known H-J equation for the generating function $S$, which in its general relativistic 
version is written as,  
\be
\frac{1}{2}\, g^{\alpha\beta}\, \frac{\partial S}{\partial x^{\alpha}}\, 
\frac{\partial S}{\partial x^{\beta}} + \frac{\partial S}{\partial \lambda} = 0
\label{HJeqn}
\en
where $\lambda$ is the geodesic affine parameter. In essence, this equation follows from the definition
of the Hamiltonian for point-particle motion,
\be
H(x^{\alpha},p_{\beta}) = \frac{1}{2}\, g^{\mu\nu}\, p_{\mu}\,p_{\nu} = -\frac{1}{2}\,\mu^2
\label{superH}
\en
by replacing the four momenta $ p^{\alpha} = dx^{\alpha}/d\lambda $ by 
$ p^{\alpha} = g^{\alpha\beta}\,\di S /\di x^{\beta}$. Here the BL coordinates
$ x^{\alpha} = \{t,r,\theta,\phi \} $ are conjugate to $ p^{\alpha} $.

One of the `miracles' of the Kerr metric is that it allows {\em full} separability of the H-J equation, leading to the 
well-known Kerr geodesic equations of motion \cite{MTW}, \cite{chandrabook}. Separability with respect 
to $\lambda,t,\phi $ generates the conserved quantities $\mu$ (the test-body's mass), $E$ (energy) and
$L_z$ (angular momentum along the symmetry axis). This is always possible as long as the
metric is stationary-axisymmetric. The special property of the Kerr metric (originating from its Petrov type-$D$ character) 
is that it allows the additional, non-trivial, separability with respect to $(r,\theta)$,
leading to the third constant $Q$, the Carter constant. This is possible {\em only} in a restricted family of coordinate frames, 
the BL one among them \cite{stewart}. In fact, if we expand the Kerr metric with respect to $a$, then the H-J equation is 
separable at each individual order.

On the other hand, the H-J equation is {\it not} fully separable with respect to the H-T metric (\ref{HT1}), not even
when the Kerr limit $\epsilon \to 0$ is taken. This reveals that the coordinate frame used in the 
original H-T metric is not `privileged' and consequently a bad choice for a quasi-Kerr metric.   
Instead, BL coordinates seem like a good choice as in the limit $\epsilon \to 0$ they 
admit separability. A similar conclusion can be reached for the general metric (\ref{mexpan}). In its present form it 
does not allow separation of the H-J equation (or of the scalar and gravitational perturbation equations) even when the
Kerr limit is taken (which corresponds to enforcing relation~(\ref{nohair})).

The H-J equation for the full quasi-Kerr metric (\ref{qKerr}) is solved by assuming the standard form,
\be
S = \frac{1}{2}\,\mu^2\, \lambda -E\,t +L_z\, \phi + {\cal S}_r(r) + {\cal S}_\theta (\theta)
\en
with an extra expansion,
\be
{\cal S}_{r,\theta} = S_{r,\theta} + \epsilon\, \hat{S}_{r,\theta}
\label{Sexpan}
\en
where $ S_{r,\theta}  $ solves the exact Kerr H-J equation. 
After separating out the Kerr portion of (\ref{HJeqn}), and neglecting ${\cal O}(\epsilon a, \epsilon^2) $ terms, we are left with,
\be
&& r\,(r-2M)\, \left [ 2\frac{dS_r}{dr}\,\frac{d\hat{S}_r}{dr} + f_3
\left (\frac{dS_r}{dr} \right )^2  \right ] + \frac{E^2\,r^3f_3}{r-2M} -K\,h_3
= -2\,\frac{dS_\theta}{d\theta}\,
\frac{d\hat{S}_\theta}{d\theta} + Z(r)\cos^2\theta\ 
\label{leftover} 
\en      
where $K$ is a separation constant and
\be
Z(r) = K h_4 -f_4\,\frac{E^2\,r^3}{r-2M} -f_4\,r\,(r-2M)\,
\left ( \frac{dS}{dr} \right )^2
\en
This non-vanishing function spoils the separability of (\ref{leftover}). In retrospect, this is not really surprising as the 
addition of the $\epsilon$-pieces `contaminates' the Kerr metric in the sense that the resulting quasi-Kerr spacetime is not 
type-$D$ anymore, although (in some loose sense) is arbitrarily `close' (this is also the explanation
for why the non type-$D$ H-T metric does not lead to a separable H-J equation).   

At this point it would seem that no real advantage has being gained by using the quasi-Kerr metric (\ref{qKerr})
as the H-J equation is still non-separable with respect to $(r,\theta)$. However, the fact that the equation is separable
in the $\epsilon \to 0$ limit can be exploited and by means of canonical perturbation theory \cite{goldstein} 
can lead to the desired equations of motion (see Section~\ref{sec:qKerr_geod}).  
        
Studying geodesic motion in a quasi-Kerr field is just the first step towards solving the EMRI problem. For gravitational wave
observations it is pivotal to be able to compute accurate waveforms and fluxes. In the case of Kerr black holes,
this is achieved by means of the Teukolsky equation \cite{teuk} (see \cite{kg_review}, \cite{chapter} for reviews
relevant to EMRI, and further references), which governs the dynamics of perturbations of the Weyl scalars
$\{\psi_0,\psi_4 \}$ in the framework of the classic Newman-Penrose formalism. The remarkable result that all
these perturbations are described by one single master equation is a consequence (once more) of the Petrov type-$D$ character 
of the Kerr metric, which allows decoupling of the various perturbation equations \cite{stewart}. 
Moreover, for a certain class of coordinate frames (the BL the most widely used one) the Teukolsky equation admits 
separation of variables, reducing the problem to the solution of simple ODEs. None of these properties survive for a 
non type-$D$ metric such as (\ref{mexpan}) or (\ref{qKerr}).
However, as the quasi-Kerr metric is `almost' type-$D$ there may still be a chance of decoupling (at least partially) 
the perturbation equations for the Weyl scalars. This crucial issue is left for future work. In the unfortunate case that
this expectation proves untrue, we may have to work with straight metric perturbations which are governed by ten coupled
PDEs plus gauge conditions. These equations will still admit separation with respect to $\phi$ so they can be
studied in the time-domain in a `$2+1$' format. In the present paper, the issue of wave dynamics will be dealt by 
simply computing approximate waveforms\footnote{A similar path was taken by Ryan who only calculated slow-motion,
`restricted Post-Newtonian', waveforms from circular equatorial orbits \cite{ryan}} (Section~\ref{sec:waveforms}),
but our conclusions should be valid even if we had rigorous waveforms at our disposal.

%%%%%%%%%%%%%%%%%%%%%%%%%%%%%%%%%%%%%%%%%%%%%%%%%%%%%%%%%%%%%%%%%%%%%%%%%%%%%%%%%%%%%%%%%%%%%

\subsection{Generic orbits}
\label{sec:qKerr_geod}

Although we are mainly concerned with equatorial orbits in this paper, we nevertheless derive approximate 
equations of motion for generic quasi-Kerr orbits. Our objective is to demonstrate how one can arrive to
such equations despite the non-separability of the H-J equation and without resorting to the full
second-order equations of geodesic motion. This subsection is somewhat detached from the paper's main
guideline and the reader should first consult Appendix~\ref{app:KerrHJ} (where we review the action-angle formalism
in the Kerr spacetime, following Ref.~\cite{wolfram}) for definitions and notation used here.

Our calculation consists of nothing more than the application of canonical perturbation theory  (see \cite{goldstein} for 
further details) to point-particle motion in the Kerr spacetime. The full Hamiltonian can be written as the sum of the Kerr 
point-particle Hamiltonian amended with the quadrupolar perturbation,
\be
H= \frac{1}{2}\, g^{\alpha\beta}_{K}\,p_{\alpha}\,p_{\beta} + \frac{1}{2}\epsilon\,h^{\alpha\beta}\,p_{\alpha}\,p_{\beta} 
\equiv H_\circ + \epsilon H_1,
\en
The unperturbed Hamiltonian is a function of the actions only, $H_\circ = H_{\circ}(I_\beta) $, while the perturbation depends on
both actions and angles, $H_1 = H_{1}(w^\alpha,I_\beta) $. Expressing the Hamiltonian in this functional form is possible
provided we have first obtained relations $x^{\kappa}(w^\alpha, I_\beta) $ and $p_{\kappa}(w^\alpha, I_\beta) $. For the Kerr problem, 
these are given by implicit relations, see Appendix~\ref{app:KerrHJ}. 

In the unperturbed problem $H_\circ $ the actions are constants and the angles vary linearly with 
respect to $\lambda $. This is no longer true for the full perturbed Hamiltonian $H$. We seek a canonical transformation 
$\{w^\alpha, I_\beta \} \Rightarrow \{ \hat{w}^\alpha, \hat{I}_\beta \} $ to a new set of angle-actions which would have 
exactly the above properties. Clearly, the old and new variables will differ at ${\cal O}(\epsilon) $. 
The generating function takes the form \cite{goldstein},
\be
F(w^\alpha, \hat{I}_\beta) = w^k \hat{I}_k + \Phi(w^\alpha,\hat{I}_\beta)
\label{Ffunc}
\en
Then we have,
\be
I_{\alpha} = \frac{\partial F}{\partial w^{\alpha}} = \hat{I}_{\alpha} + \frac{\partial \Phi}{\partial w^{\alpha}} \qquad
\mbox{and} \qquad \hat{w}^{\alpha} = \frac{\partial F}{\partial \hat{I}_{\alpha}}   = 
w^{\alpha} + \frac{\partial \Phi}{\partial \hat{I}_{\alpha}}
\label{oldnew}
\en
Denoting as $K(\hat{I}_\beta,\epsilon) $ the new Hamiltonian we also have,
\bea
\frac{d\hat{w}^{\alpha}}{d\lambda} &=& \frac{\partial K}{\partial \hat{I}_\alpha} \quad \Rightarrow \quad
\hat{w}^\alpha = \hat{\nu}_\alpha \,\lambda + \beta^\alpha
\\
\frac{d\hat{I}_\alpha}{d\lambda} &=& -\frac{\partial K}{\partial \hat{w}^\alpha} = 0 + {\cal O}(\epsilon^2)   
\ena
where $\hat{\nu}_\alpha $ are the new fundamental frequencies. 
The function $F$ solves the following H-J equation 
\be
H(w_\alpha,\frac{\partial F }{\partial w^\alpha}) = K(\hat{I}_\beta,\epsilon)
\label{newHJ}
\en
The standard choice for the Hamiltonian K is,
\be
K(\hat{I}_\beta,\epsilon) = H_{\circ}(\hat{I}_\beta) + \epsilon\, \langle H_{1} \rangle (\hat{I}_\beta)
\en
where we have defined the {\it averaged} perturbation with respect to the original angle variables,
\be
\langle H_{1} \rangle(\hat{I}_{\beta}) \equiv \oint dw^\alpha H_{1}(w^k,\hat{I}_\beta)
\en
Expanding the left-hand side of (\ref{newHJ}) we then find for the generating function,
\be
\nu_\alpha \frac{\partial \Phi}{\partial w^\alpha} = \epsilon \left [\langle H_{1} \rangle
-H_{1}  \right ]
\label{GenPhi}
\en

The perturbed equations of motion are derived by first inverting eqns.~(\ref{oldnew}) with respect to the old action-angles,
isolating the secular changes in these relations and then substituting in the unperturbed equations $w^{\alpha}(x^k,I_\beta) $ 
(see eqns.~(\ref{angle_Kerr2}) in Appendix~\ref{app:KerrHJ} ). Note that in the function $\Phi$ we are free to interchange
new with old actions at ${\cal O}(\epsilon) $ accuracy. From (\ref{GenPhi}) and the fact that $\Phi(w^\alpha, I_\beta) $ is
a periodic function of the angles $w^\alpha $  \cite{goldstein} we find that,
\be
 \langle \frac{\partial\Phi}{\partial w^\alpha} \rangle = 0
\en 
which implies that there are no secular changes in the actions at ${\cal O}(\epsilon)$. This also means that there is no
secular change in $\{ E,L_{z},Q \} $. At leading order, quasi-Kerr orbits admit three integrals of motion, like in the Kerr
geodesic motion.  

Averaging the perturbation in the angle variables 
$ \langle \partial \Phi /\partial I_\alpha (w^k,I_\beta) \rangle = \langle \partial \Phi /\partial I_\alpha \rangle (I_\beta) \neq 0$   
implies that in a secular sense  $w^\alpha = \hat{\nu}\, \lambda + \hat{\beta}^\alpha $ where $\hat{\beta}^\alpha $ a constant phase. 

Therefore we have shown that the quasi-Kerr geodesic equations of motion can be directly derived from the corresponding 
Kerr equations (\ref{angle_Kerr2}) with the substitution $\nu_\alpha \Rightarrow \hat{\nu}_\alpha $ {\em in the left-hand
sides only }. 
Despite the identical shapes between Kerr and quasi-Kerr orbits, the difference in the frequencies will translate in
a difference in precession rates for the periastron and for the orbital plane \footnote{A basic feature of
generic Kerr orbits is their characterisation by three {\it incommensurate } orbital frequencies
$\Omega_i = \nu_i/\nu_t $ (in BL time). This property results in orbits which are not closed, but rather
exhibit precessional motions. Periastron precession is a consequence of $\Omega_r \neq \Omega_\phi $ 
and orbital plane precession (Lense-Thirring precession) is a consequence of $\Omega_\theta \neq \Omega_\phi $.
Keplerian orbits, on the other hand, are fixed closed ellipses since their frequencies are degenerate, 
$\Omega_r =\Omega_\theta=\Omega_\phi $. }.    

Generic quasi-Kerr orbits will be discussed in detail in a future paper. Here we have shown that in a secular
sense the motion has three conserved quantities (apart from $\mu$), which is an important property and stands above
of what one would expect for a general non-Kerr axisymmetric-stationary metric like (\ref{mexpan}). As we mentioned in the 
Introduction, geodesic motion in such metric is not formally integrable and may well lead to chaotic behaviour in a test-body motion. 

In fact, orbital dynamics in general axisymmetric-stationary gravitational fields are typically much different than the ones in 
spherically-symmetric fields or even the Kerr field. We can easily demonstrate this difference, by asking the simple question: which
axisymmetric-stationary gravitational fields can support circular orbits? In Newtonian theory the answer is that only a special 
family of potentials with the form $V(r,\theta) = V_\circ(r) + V_1(\theta)/r^2 $ (where $V_\circ, V_1$ arbitrary functions) 
admit orbits with $u^r= du^r/dt = 0 $. In General Relativity the same requirement for a given metric $g_{ab}(r,\theta) $ results in 
conditions involving the metric and its first derivatives. Once more, the Kerr metric is special as it satisfies these conditions. 
Further analysis of this issue will be presented in a future paper.

%%%%%%%%%%%%%%%%%%%%%%%%%%%%%%%%%%%%%%%%%%%%%%%%%%%%%%%%%%%%%%%%%%%%%%%%%%%%%%%%%%%%%%%%%%%%%%%

\subsection{Equatorial orbits}
\label{sec:equat}

For the special case of equatorial orbits the quasi-Kerr H-J equation~(\ref{HJeqn}) 
is trivially separable. The resulting equation of motion are,
\bea
(u^r)^2 = \left(\frac{dr}{d\lambda}\right)^2 &=& V_r(r) = 
(E^2 -1) + \frac{2M}{r} - [L_z^2 + a^2(1-E^2)]\frac{1}{r^2} + \frac{2M}{r^3}(L_z -aE)^2 - 
\nonumber \\
& & \epsilon\left(1-\frac{2M}{r}\right)\left[ (f_3-h_3)\frac{L_{z}^2}{r^2} + f_3 \right]
\\
u^\phi = \frac{d\phi}{d\lambda} &=& V_{\phi} =  \frac{1}{\Delta}\left[ \frac{2M}{r}(aE-L_z)+L_z \right] -
\epsilon\frac{h_3}{r^2} L_z 
\\
u^t = \frac{dt}{d\lambda} &=& V_t = \frac1{\Delta}\left[ E(r^2+a^2) + \frac{2Ma}{r}(Ea-L_z)\right]-
\epsilon\left(1-\frac{2M}{r}\right)^{-1}f_3E
\label{geod_equat}
\ena
In the standard manner, we express the test-body's radial location in terms of a pair of orbital
elements, the semi-latus rectum $p$ and eccentricity $e$,
\be
r = \frac{p}{1+e\cos{\chi}}
\en
In the weak-field limit ($p\gg M$) these parameters coincide with the familiar Keplerian parameters.   
The parameter $\chi$ varies monotonically, while $r$ `runs' between the two radial turning
points, the periastron $r_p = p/(1+e)$ and the apastron $r_a = p/(1-e) $.  
The elements $(p,e)$ can be written as functions of $E,L_z$ (and vice-versa) using $V_r(r_a)=V_r(r_p)=0 $. 
For the Kerr orbits $V_r(r) $ is a cubic polynomial and the third 
root $r_3$ (which is the smallest of the three) can be used to define a `separatrix' of bound/unbound orbits. 
The transition occurs when $r_p=r_3 $ which defines an innermost stable bound orbit. In the quasi-Kerr case, the 
equation $V_r =0$ is more complicated, but since the equations of motion (\ref{geod_equat}) deviate only slightly from the 
corresponding Kerr equations, we expect the quasi-Kerr separatrix to be located close to the Kerr separatrix \cite{kgdk} 
for the same $\{a,p,e\}$. 

Equatorial orbits are characterised by a pair of orbital frequencies $\Omega_r, \Omega_\phi$. These are defined
in terms of the radial period $T_r$ (the coordinate time required for the body to move from $r_p $ to $r_a$ and back),
and of the periastron shift $\Delta\phi $ (total accumulated angle $\phi$ over a period $T_r$),
\be
\Omega_r = \frac{2\pi}{T_r}, \quad \Omega_\phi = \frac{\Delta\phi}{T_r}
\label{freqs}
\en
These frequencies are `observables' as the emitted gravitational waveform (on an adiabatic timescale where
backreaction is not significant) comes in discrete frequencies, 
\be
\omega_{mk} = m\,\Omega_\phi + k\,\Omega_r
\label{spectrum}
\en   
where $k,m$ are integers (the latter associated with the azimuthal angle $\phi$). The appearance of a spectrum
of the form (\ref{spectrum}) is a direct consequence of the periodicity of the motion. Note that in the Keplerian
limit ($p\gg M$) we have $\Delta\phi \to 2\pi $, hence $ \Omega_r \to \Omega_\phi $. 

\begin{figure}[ht]
\includegraphics[height=8cm,width=14cm]{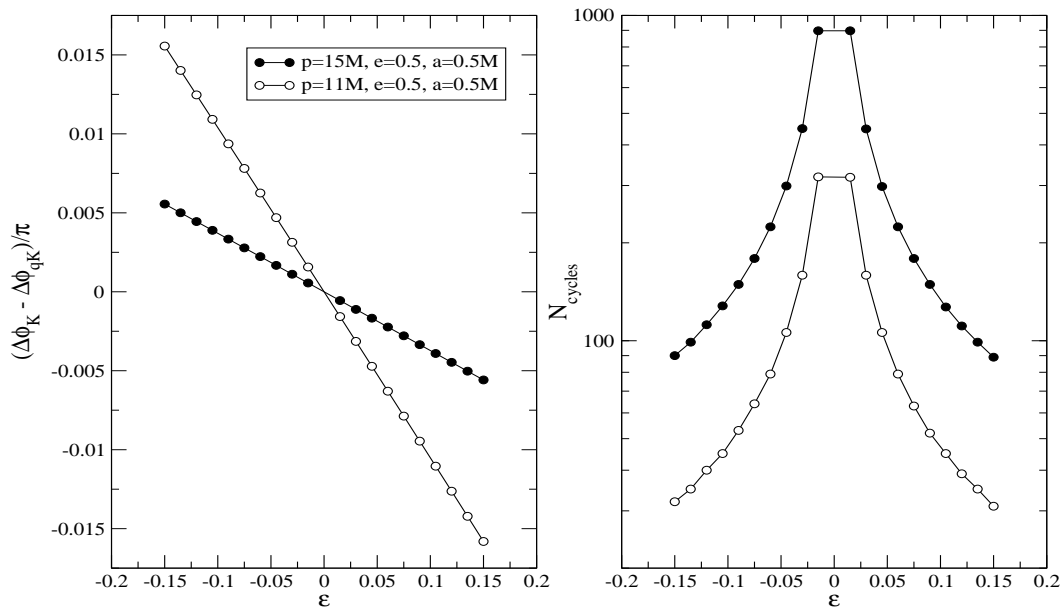}
\caption{Left panel: periastron shift difference $\Delta\phi_{\rm K} - \Delta\phi_{\rm qK}  $ as a function of 
the deviation $\epsilon$. Right panel: number of cycles ${\cal N} $ required to accumulate $\pi/2$ 
difference in periastron shifts (defined by eqn.(\ref{Ncycl})) as function of $\epsilon$. 
For both panels, we have considered two orbits: $a=0.5M, e=0.5$ and $p=10M,\, p=15M$.}
\label{fig:periastron}
\end{figure}

\begin{figure}[ht]
\includegraphics[height=8cm,width=12cm]{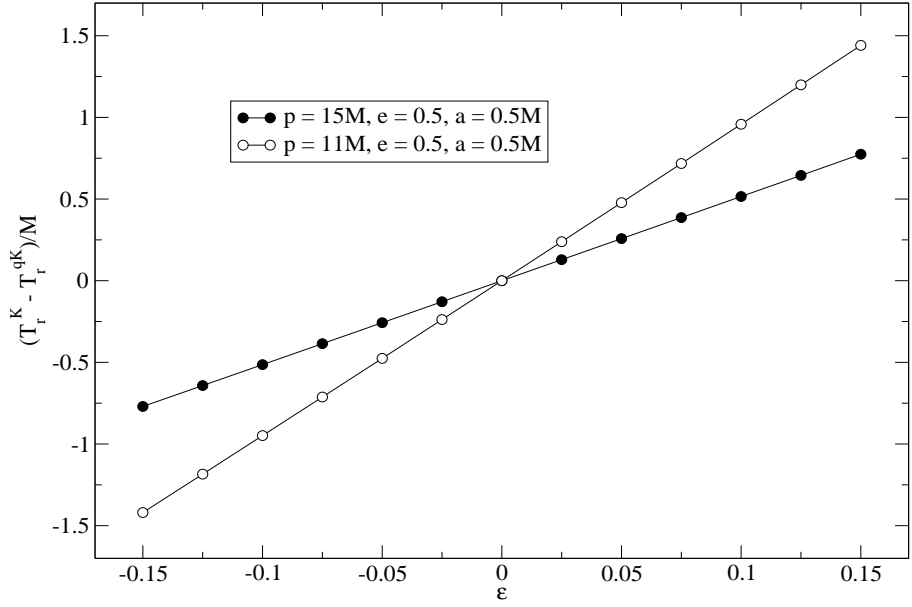}
\caption{Difference in the radial period $T_r $ Kerr and quasi-Kerr orbits with $a=0.5M, e=0.5$ and $p=10M,\, 
p=15M$.}
\label{fig:periods}
\end{figure}

\begin{figure}[ht]
\includegraphics[height=8cm,width=14cm]{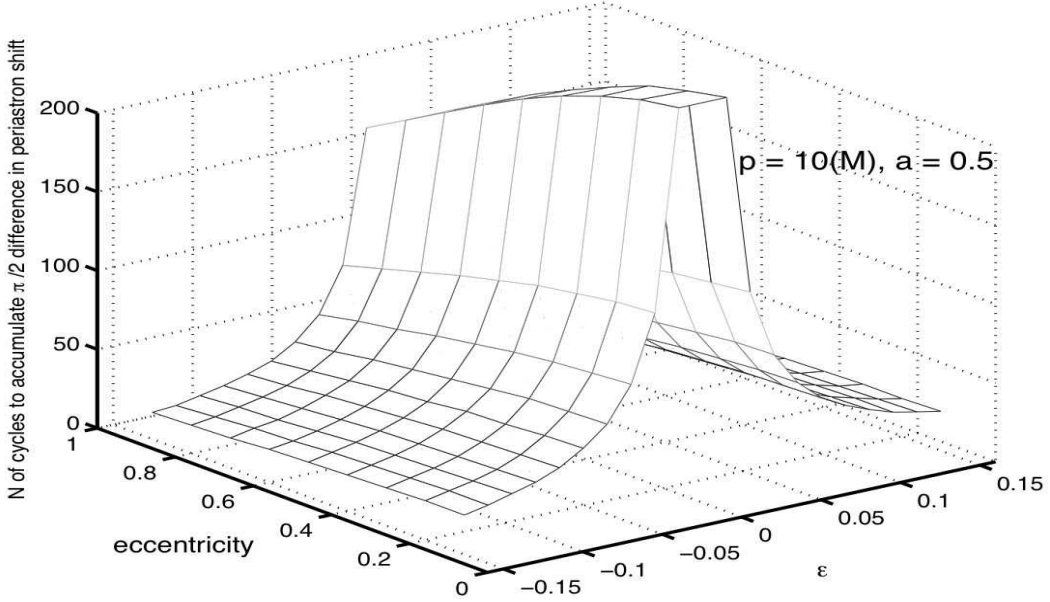}
\caption{Number of cycles to accumulate $\pi/2$ difference between
Kerr and quasi-Kerr periastron shifts as a function of $\epsilon$ and eccentricity $e$.
We have fixed the other two parameters: $p=10M$, $a=0.5M$.} 
\label{fig:cycles}
\end{figure}

As a first stab at the problem of comparing Kerr vs quasi-Kerr EMRIs we calculate the above frequencies
$\Omega_r,\Omega_\phi $ for {\it identical} values $\{a,p,e\}$ while varying the quadrupole deviation $\epsilon$. 
We use the simple expressions,
\be
T_r = \int_{r_p}^{r_a}dt\,\frac{u^t}{u^r} = \int_{0}^{2\pi}d\chi\,\frac{dr}{d\chi}\frac{u^t}{u^r}
\en  
and
\be
\Delta\phi = \int_{0}^{T_r}dt\,\frac{u^\phi}{u^t} = \int_{0}^{2\pi}d\chi \frac{dr}{d\chi}\,\frac{u^\phi}{u^r}
= \int_0^{2\pi}d\chi\, \frac{ep\sin{\chi}} {(1+e\cos{\chi})^2}\frac{V_{\phi}(r(\chi))}{\sqrt{V_r(r(\chi))}}
\en
Results are shown in Figs.~\ref{fig:periastron},~\ref{fig:periods} \& \ref{fig:cycles} 
for the pair of orbits $(p,e) = (10M, 0.5), (15M, 0.5) $ and $a=0.5M $. A quantity which nicely illustrates the difference  
between Kerr and quasi-Kerr orbits is the number of cycles ${\cal N}$ required to accumulate $\pi/2$ difference in periastron shift:
\be
\mathcal{N} = \frac{\pi/2}{|\Delta\phi_{\rm K} - \Delta\phi_{\rm qK}|}
\label{Ncycl}
\en 
In Fig.~\ref{fig:cycles} we show how ${\cal N} $ varies with $\epsilon$ and eccentricity $e$ for fixed $p$ and $a$. 
The dependence on $e$ is quite weak, but there is a strong dependence on $\epsilon$. For example, for 
a moderate value $|\epsilon| = 0.05 $ which corresponds to a fractional difference $\sim 8\% $ in the quadrupole
moment, we only need about $\sim 100-200 $ orbits to accumulate $\pi/2 $ difference in $\Delta\phi $. Such large
orbital dephasing would certainly manifest itself in the waveforms as we shall see in the following Section.

%%%%%%%%%%%%%%%%%%%%%%%%%%%%%%%%%%%%%%%%%%%%%%%%%%%%%%%%%%%%%%%%%%%%%%%%%%%%%%

\section{Waveform comparisons}
\label{sec:waveforms}

\subsection{Kerr vs quasi-Kerr}
\label{sec:KvsqK}

As we already pointed out in Section~\ref{sec:separability}, the calculation of rigorous 
gravitational waveforms and fluxes is a major challenge for any non-Kerr EMRI program. 
For the purposes of this study, which is just an initial investigation, it will suffice to
use approximate waveforms, in particular the `hybrid' waveforms discussed in \cite{kg_review},\cite{kludge_paper}. These waveforms are 
generated using flat spacetime wave formulae (see \cite{MTW},\cite{press} for more details) for the two components 
$\{ h_{\times},h_{+} \}$, coupled with exact relativistic geodesic motion. At least for the particular case of Kerr, 
it has been established that these waveforms agree quite well with rigorous Teukolsky-based waveforms \cite{kludge_paper}. 
Hence, it makes sense to assume that the same will be true for the quasi-Kerr case.

In this Section we shall compare quasi-Kerr against Kerr waveforms with identical orbital
parameters $(p,e)$, spin $a/M $ and for the same kinematic initial conditions. Since a nonzero $\epsilon $ imparts
a change in the orbital frequencies we would expect this change to manifest itself as a significant cumulative phase 
difference between the corresponding waveforms after several orbits. This difference can be rigorously quantified in terms of 
the overlap function, that is, the scalar product between the normalised waveforms within the LISA sensitivity band:
\be
(\hat{h}_1,\, \hat{h}_2) = 4 \Re \int_0^{\infty}  
\frac{\tilde{h}_1(f)\tilde{h}_2^*(f) }{S_h(f)}\; df
%\sum_{k=0}^{N-1}\frac{\tilde{h}_1(f_k)
%\tilde{h}_2^*(f_k) + \tilde{h}_2(f_k)\tilde{h}_1^*(f_k)}{S_h(f_k)}
%\Delta f,
\label{olap}
\en
where a tilde denotes the Fourier transform of the waveform, a star stands for the complex conjugate, a hat 
denotes normalised waveform $h(t_k)$ according to $(\hat{h}_1,\hat{h}_1) =(\hat{h}_2,\hat{h}_2) = 1$ and $S_h(f)$ 
is the expected LISA sensitivity function (one-sided power spectral density).
To calculate this latter function we use an analytic fit (discussed in  
Ref.~\cite{kludge_paper}) to the numerically-generated sensitivity curve of Ref.~\cite{LISAcurve}, for one year of 
observation and omitting any confusion noise\footnote{We have observed that addition of confusion noise does not affect the 
overlaps, although it slightly affects the estimated radiation reaction time scale.}.

The overlap defined in (\ref{olap}) is closely related to the signal-to-noise ratio (SNR) \cite{CF}, which 
is given by, 
$$
SNR = (\hat{h},\,s) = \mathcal{A}(\hat{h},\,\hat{h}) = \mathcal{A} =
(s,s)^{1/2}
$$
Here we have assumed that the template $\hat{h}$ exactly matches the true signal $s$, and $\mathcal{A}$ is the amplitude of the 
signal in units of the normalised waveform $\hat{h}$. An imperfect template or a disagreement between the template's 
parameters and those of the signal results in a reduced SNR,
\bea
SNR = (\hat{h}_1,\,s) = \mathcal{A}(\hat{h}_1,\,\hat{h}_2) < 
\mathcal{A}
\label{SNR}
\ena
In this expression we have assumed that the normalised template $\hat{h}_1$ does not exactly match the signal $s= 
\mathcal{A} \hat{h}_2$. In other words, a low overlap between a template and the expected signal implies a drop in SNR (which is
accompanied by a drop in the event rate due to the decrease of the observable spatial volume). 

\begin{figure}[ht]
\includegraphics[height=8cm,width=14cm]{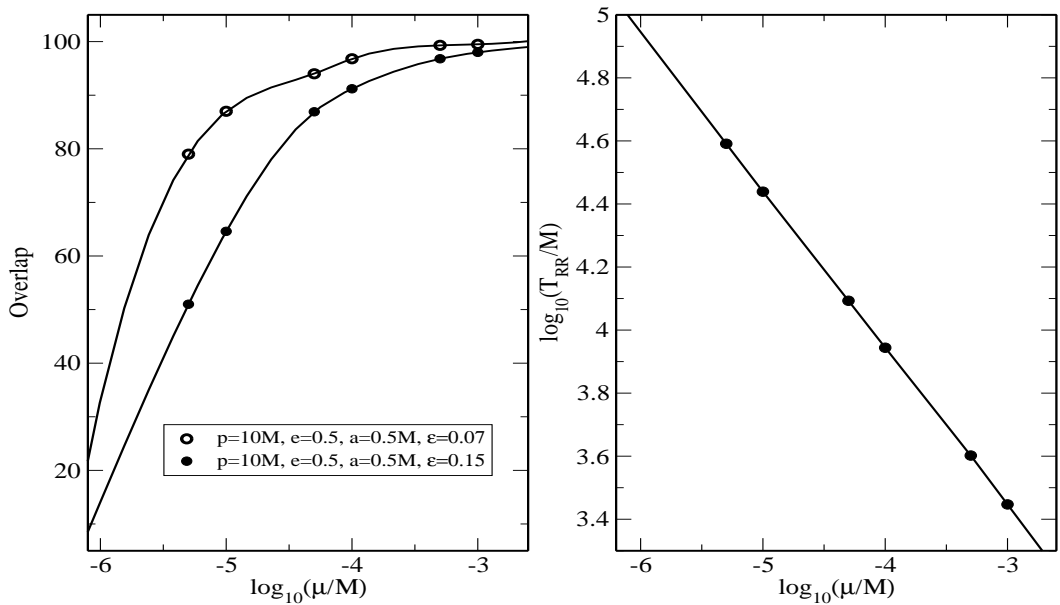}
\caption{Right panel: radiation reaction timescale $T_{\rm RR} $ (see main text for definition),
as a function of mass ratio $\mu/M $. 
Left panel: overlaps (expressed in $\%$) between quasi-Kerr and Kerr waveforms for the same orbit
$p=10 M, e=0.5, a = 0.5 M $ and for $\epsilon = 0.07, 0.15 $, truncated at $t=T_{\rm RR}(\mu/M)$. The solid line is the spline 
interpolation between the data points.}
\label{fig:Olap}
\end{figure}

Since we neglect any radiation reaction effect on the orbital motion our calculations will be consistent
provided we `truncate' any waveform at the radiation reaction time scale $T_{\rm RR}$. 
We define $T_{\rm RR}$ according to the following rule. For a test-body in Kerr spacetime we can include orbital backreaction
using the approximate hybrid method \cite{kg_review}, \cite{GHK}, \cite{GG}. Then comparing Kerr waveforms with and without 
radiation reaction (denoted as $h_K(t),h_K^{RR}(t) $, respectively) we can define $T_{\rm RR}$ as the truncation time $T_{tr}$
at which the overlap between these two waveforms drops below $95\% $. 
\be
\left( \hat{h}_{\rm K}(t = T_{\rm RR}),\, \hat{h}^{\rm RR}_{\rm K}(t= T_{\rm RR}) \right) = 0.95
\label{TRR}
\en
The particular threshold 0.95 was chosen as it coincides with the usual value of `minimal match' for constructing template 
banks \cite{CF}, \cite{Bala}. Based on our definition for $T_{\rm RR}$ we can safely approximate the small body's motion as a
geodesic for a time interval $ t \lesssim T_{\rm RR} $. We have numerically computed $T_{\rm RR} $ (see Fig.~\ref{fig:Olap})
as a function of $\mu/M $, and not surprisingly, we find a linear dependence, $ T_{\rm RR} \sim M/\mu $.

\begin{figure}[ht]
\includegraphics[height=8cm, width=15cm]{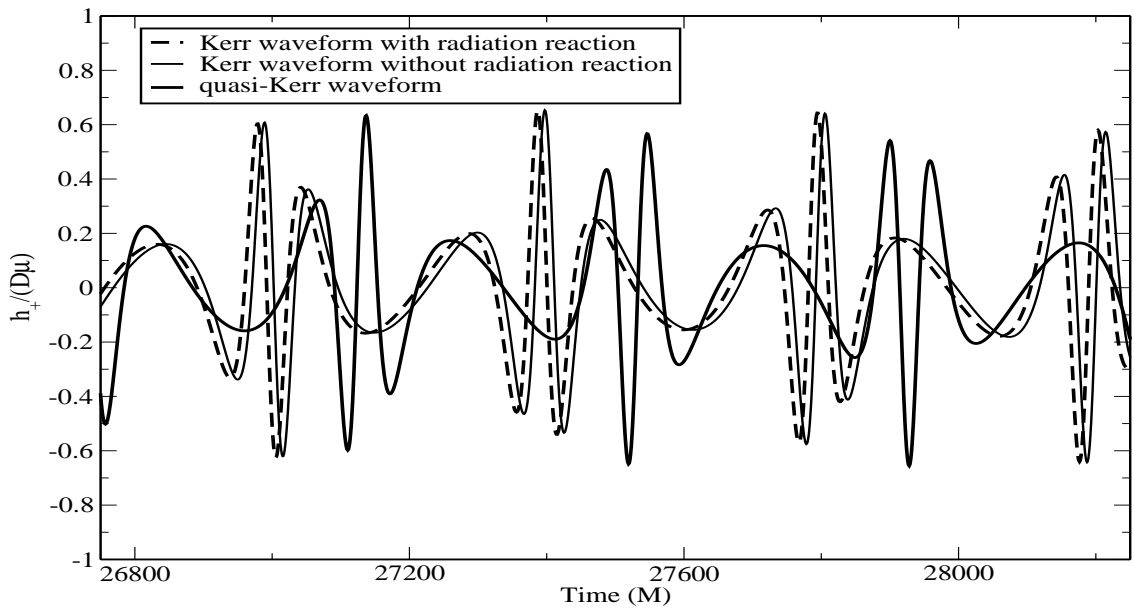}
\caption{Comparing quasi-Kerr (solid curve) and Kerr (dashed curve) approximate hybrid waveforms 
for the orbit $p= 10 M, e=0.5, a=0.5 M$ and for $\epsilon = 0.15$. In addition, we include a Kerr waveform taking
into account backreaction on the orbit (assuming the same initial orbit), represented by the dashed curve. 
All waveforms are shown at a time window close to the radiation reaction timescale $T_{\rm RR}$.}  
\label{fig:wvf} 
\end{figure}

For the purpose of waveform comparison we select the moderately relativistic orbit
$p= 10M, \, e=0.5,\, a=0.5$. For a mass ratio $\mu/M \sim 10^{-5} $ we have $T_{\rm RR} \approx 27500 M$. 
The quadrupolar deviation is fixed at $\epsilon = 0.15$ which corresponds to a fractional difference
$\delta \cQ/\cQ^{\rm K} = 40\% $. The actual waveforms are shown in Fig.~\ref{fig:wvf}. Clearly, after a time lapse 
$\sim T_{\rm RR} $ the accumulated phase-difference between the two waveforms is quite significant. The waveform deviation 
due to a nonzero $\epsilon $ can be also compared against the deviation imparted by allowing the orbit to evolve under 
radiation reaction. For the particular case of Fig.~\ref{fig:wvf}, after a time interval $t\sim T_{\rm RR}$, the waveforms 
dephase mainly due to the non-Kerr property, rather than due to radiation reaction.

In order to quantify the waveform difference in a more rigorous manner we compute overlaps between Kerr and quasi-Kerr waveforms 
truncated at $t=T_{\rm RR}$. Some results are given in Fig.~\ref{fig:Olap} as a function of the mass ratio for the two values
$\epsilon = 0.07$ and $\epsilon = 0.15$. One can see that the overlap can drop quite dramatically (down to 40\%)
even for small values of $\epsilon$ (0.07) and for realistic mass ratios ($\sim 10^{-6}$). 

This an important result of the present paper: The typically low overlaps between quasi-Kerr and Kerr waveforms 
(for the same orbital parameters and spin) simply mean that using Kerr templates for studying EMRIs in a true non-Kerr spacetime might 
result in low SNRs, accompanied by significant loss in the number of observed events.

%%%%%%%%%%%%%%%%%%%%%%%%%%%%%%%%%%%%%%%%%%%%%%%%%%%%%%%%%%%%%%%%%%

\subsection{Prelude to the `confusion problem'}
\label{sec:confusion}

According to our results, it is clear that even a modest (say $\sim 10\% $) deviation from the Kerr quadrupole
moment is sufficient to render any Kerr template waveform inaccurate assuming the {\it same} orbit
$\{p,e\} $ and spin parameter $a/M$. 

\begin{figure}[ht]
\includegraphics[height=8cm,width=10cm]{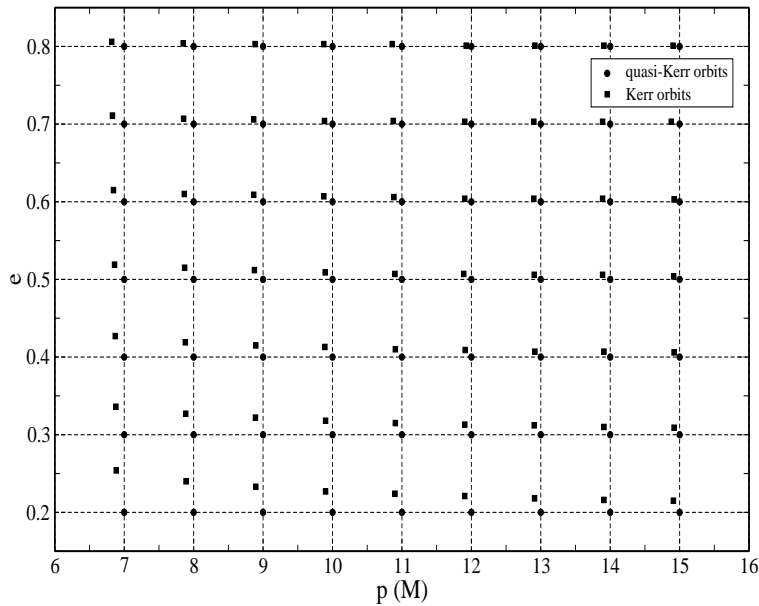}
\caption{Deviation in the orbital elements $p,e$ between Kerr (squares) and quasi-Kerr (circles), after equating
the orbital frequencies $\Omega_r, \Omega_\phi $, and for fixed $ a=0.4M, \epsilon=0.1 $. }
\label{fig:grid}
\end{figure}

This statement begs the question: is it actually best to compare waveforms corresponding to the same
set of parameters $\{a,p,e \}$ ? One could argue that we should have rather fixed the orbital frequencies,
\be
\Omega_i^{\rm K} = \Omega_i^{\rm qK}, \qquad i= \phi,r
\label{sameOm}
\en 
Then one would expect a good agreement in phase between the two waveforms. Note that this choice would correspond in 
comparing waveforms with different $\{p,e \} $, given fixed values of $a,\epsilon $. If $(p,e)$ and $(\tilde{p},\tilde{e}) $ 
are the Kerr and quasi-Kerr orbital elements, respectively, and assuming a small deviation 
$\delta p = \tilde{p} -p \sim {\cal O}(\epsilon)$ (and similarly for $e$) we have,  
\bea
&& \Omega_i^{\rm K}(p,e) =\Omega_i^{\rm qK}(\tilde{p},\tilde{e}) = \Omega_i^{\rm K}(\tilde{p},\tilde{e}) + 
\epsilon\,\Omega^{(1)}_i(p,e) \Rightarrow 
\nonumber \\
\nonumber \\
&&\frac{\partial \Omega_i^{\rm K}}{\partial p}\,\delta p + \frac{\partial \Omega_i^{\rm K}}{\partial e}\,\delta e
= -\epsilon\,\Omega^{(1)}(p,e)
\label{dpde}
\ena
This  $2\times 2 $ system has a non-vanishing determinant and a unique solution for $(\delta p,\delta e) $. Note that for 
$\epsilon =0 $ this solution becomes trivial, but this is true as long as we do not allow for variation in $a/M $.
In Fig.~\ref{fig:grid} we show the solution of (\ref{dpde}) for $a=0.4M,~\epsilon=0.1 $. We can see that the most pronounced
deviation between Kerr and quasi-Kerr orbits occurs in the small $p$, small $e$ region. This is understandable as in this
case the body spends most of its orbital time in the strong-field where the deviation from the Kerr metric is stronger.   
As expected for $p\gg M$ both $ \delta p,~\delta e \to 0 $.     

Our preliminary results on waveform comparison are quite alarming: fixing the orbital frequencies instead of $\{p,e\}$ 
lead to overlaps very close to unity  between Kerr and quasi-Kerr waveforms. For the particular case of $\mu/M = 10^{-5}$, $a= 0.3M$ 
and $\epsilon = 0.1$ and for $\{\tilde{p},\tilde{e} \} = \{10M,0.3 \}$ we get from (\ref{dpde}) 
$\{p,e \}_{K} = \{9.906M,0.317 \} $ and the resulting overlap is $97.6\% $ (for the time interval $T_{\rm RR}\approx 31000 M $).  
Hence, we have naturally stumbled upon a case of the {\it confusion problem}: the possibility that for a given non-Kerr EMRI 
waveform there might be a corresponding Kerr waveform with a different set of parameters which could mimic the former 
(that is produce an overlap $\geq 95 \% $)  on a radiation reaction timescale. 

Here we have given one example where this scenario is indeed true, but this is not the end of the story. Allowing a time window, 
longer than $T_{\rm RR} $, for which radiation reaction effects become important the overlap could significantly reduce as a 
consequence of different orbital evolution between Kerr and quasi-Kerr. Orbits which are initially `tuned' to have almost 
identical orbital frequencies are not expected to preserve this property for $ t \gtrsim T_{\rm RR} $.    

The non-trivial issue of waveform confusion is of great importance for LISA and certainly requires further investigation. We will 
address it in more detail in a following paper.

%%%%%%%%%%%%%%%%%%%%%%%%%%%%%%%%%%%%%%%%%%%%%%%%%%%%%%%%%%%%%%%%%

\section{Concluding discussion}
\label{sec:conclusions}

One of the LISA mission main science goals will be the `mapping' of Kerr spacetime. In order to perform this crucial test of General
Relativity we should be in a position to prepare data analysis tools capable of gauging any possible deviations from the
Kerr metric. In this paper we have introduced a practical scheme which could be a strong candidate for this task.
Our scheme is based on the notion of a `quasi-Kerr' metric, that is, a metric that deviates only slightly from the known Kerr metric, 
while still being stationary and axisymmetric. This deviation can first appear in the value of the spacetime's quadrupole moment. We only 
consider such deviations and ignore any other possible disagreement in higher multipole moments. 

Our quasi-Kerr metric is built with the help of the exterior Hartle-Thorne metric, which describes the spacetime outside any 
slowly-rotating, stationary axisymmetric body. Given this metric, we have first studied geodesic motion of a test-body. 
Despite the fact that the Hamilton-Jacobi equation is non-separable, we manage to derive approximate equations for generic orbits 
employing canonical perturbation theory. 

In this paper we mostly focused on equatorial orbits (in which case the Hamilton-Jacobi equation is trivially separable) 
and for given orbital parameters and spin we computed the change imparted to the orbital frequencies and periastron advance by the non-Kerr 
metric deviation. We found that for (say) a $\sim 10\% $ fractional difference in the quadrupole moment, the cumulative change in these 
quantities is significant after about $\sim 100$ orbits, and even for moderately relativistic orbits, it significantly affects the 
gravitational wave phasing. This was demonstrated by generating approximate `hybrid' waveforms (for orbits with identical $p,e$), for 
both Kerr and quasi-Kerr spacetimes, and computing their mutual overlap function. 

These overlaps typically drop below $\sim 60 \% $ for the relevant mass ratio (allowing for a time interval at which any radiation
reaction effects are still not important). Such low overlaps mean that we might expect significant loss in signal-to-noise ratio if 
we attempt to use Kerr waveform templates for the detection of a gravitational wave signal generated by an EMRI in a quasi-Kerr field. 
This statement holds as long as we compare waveforms with the same orbital elements. However, the picture becomes more blurred
when we compare waveforms which correspond to different orbits but same orbital frequencies. Then it is possible for a pair of
Kerr and quasi-Kerr waveforms to match extremely well. This confusion problem could be a generic feature of any non-Kerr data analysis
effort, and it means that we may confuse a possible true non-Kerr event with a Kerr EMRI with the wrong parameters. A possible way to break 
this degeneracy is to include backreaction effects, in which case orbits with initial identical frequencies 
diverge on the radiation-reaction timescale. This issue will be further analysed in a following paper. Similarly, we shall 
extend the present calculations to the case of non-equatorial orbits, but we expect that the conclusions of the present paper will 
remain true in this more general scenario.  

The scheme we propose here clearly requires further development, especially towards the direction of computing rigorous 
quasi-Kerr inspiral waveforms. The subject of wave-dynamics in a general vacuum axisymmetric-stationary spacetime or even
a less general spacetime like the quasi-Kerr, is almost unexplored. Work in progress aims to investigate if it would be useful to work 
within the Newman-Penrose formalism \cite{chandrabook}, trying to formulate some approximate `quasi-Teukolsky' equation, 
or if it would be more practical to work directly with metric perturbations.

\acknowledgements

KG thanks Nils Andersson, Leor Barack, John Miller and James Vickers for very fruitful discussions and 
acknowledges support from PPARC Grant PPA/G/S/2002/00038. SB would like to thank Luc Blanchet for 
useful comments during the GWDAW9 meeting.

%%%%%%%%%%%%%%%%%%%%%%%%%%%%%%%%%%%%%%%%%%%%%%%%%%%%%%%%%%%%%%%%%%%%%%%%%

\appendix

\section{}
\label{app:HT_funcs}

Here we provide the explicit form of the various functions appearing in the H-T metric, eqn.~(\ref{HT1}). 
\bea
F_1(r,\theta) &=&  -\frac{1}{8Mr^4(r-2M)} [\, (r-M)(16 M^5 + 8M^4 r
-10 M^2 r^3 -30 M r^4 + 15 r^5)
\nonumber \\
\nonumber \\
&+& \cos^2\theta (48M^6 -8M^5 r -24 M^4 r^2 - 30 M^3 r^3 
-60 M^2 r^4 + 135 M r^5 -45 r^6) ] + A_1(r,\theta)
\nonumber \\
\nonumber \\
&\equiv & f_1(r) + f_2(r)\, \cos^2\theta \,,
\\
\nonumber \\
F_2(r,\theta) &=& - \frac{5(r-M)(2M^2 +6Mr -3r^2)}{8Mr(r-2M)} 
\{ 1- 3 \cos^2\theta \} -A_1(r,\theta) 
\nonumber \\
\nonumber \\ 
&\equiv& f_3(r) + f_4(r)\, \cos^2\theta \,,
\\
\nonumber \\
G_1(r,\theta) &=& \frac{(P-72M^5 r) -3\cos^2\theta (L-56M^5 r)}{8Mr^4(r-2M)}
- A_1(r,\theta) \quad \equiv g_1(r) + g_2(r) \cos^2\theta\,, 
\\
\nonumber \\
H_1(r,\theta) &=& \frac{(16 M^5 + 8M^4 r -10 M^2 r^3 +15Mr^4 + 15r^5)}{8Mr^4}
\{ 1 -3 \cos^2\theta \} + A_2(r,\theta) 
\nonumber \\
\nonumber \\
&\equiv& h_1(r) + h_2(r) \cos^2\theta \,,
\\
\nonumber \\
H_2(r,\theta) &=& \frac{5(2M^2 -3Mr -3r^2)}{8Mr} \{ 1-3\cos^2\theta \} 
-A_2(r,\theta) \quad \equiv h_3(r) + h_4(r) \cos^2 \theta \,,
\ena
where 
\bea
P(r) &=& 15\, r^6 -45\,M r^5 + 20\,M^2 r^4 + 10\, M^3 r^3 + 8\, M^4\, r^2 + 80\, M^6 
\\
\nonumber \\
A_1(r,\theta) &=& \frac{15r(r-2M)}{16M^2} \ln \left (\frac{r}{r-2M}  
\right ) \{ 1 -3\cos^2\theta \}\,,
\\
\nonumber \\
A_2(r,\theta) &=& -\frac{15(r^2 -2M^2)}{16M^2} \ln \left ( \frac{r}{r-2M} 
\right ) \{ 1- 3 \cos^2 \theta \}\,.
\ena

The radial functions ${\cal F}_{1,2}(r) $ appearing in the quasi-Kerr metric (\ref{qmetric}) are:
\bea
{\cal F}_{1}(r) &=& -\frac{5(r-M)}{8Mr(r-2M)}\,(2M^2 + 6Mr -3r^2) -\frac{15 r(r-2M)}{16 M^2}\,\ln\left ( \frac{r}{r-2M} \right )
\\
\nonumber \\
{\cal F}_{2}(r) &=& \frac{5}{8Mr}\,(2M^2 -3Mr -3r^2) + \frac{15}{16M^2} (r^2 -2M^2)\ln \left ( \frac{r}{r-2M}\right )  
\ena

%%%%%%%%%%%%%%%%%%%%%%%%%%%%%%%%%%%%%%%%%%%%%%%%%%%%%%%%%%%%%%%%%%%%%%%%%%%%%%

\section{}
\label{app:KerrHJ}

In this Appendix we present, in some detail, the Hamilton-Jacoby/action-angle theory for a test-body
in the Kerr spacetime, see Refs.~\cite{MTW}, \cite{wolfram}, \cite{carter2} for full exposition to the topic. 

The starting point is the Hamiltonian for point-particle motion ($\alpha,\beta, \kappa \in \{t,r,\theta,\phi \} $),
\be
H_\circ(x^{\alpha},p_{\beta}) = \frac{1}{2}\,g^{\alpha \beta}_{\rm K}\,p_{\alpha}\,p_{\beta} =-\frac{1}{2}\mu^2 
\en 
Generating a canonical transformation $\{ x^{\alpha}, p_{\beta}  \} 
\Rightarrow \{ Q^{\alpha}, \gamma_{\beta}\}$,
with the requirement of a vanishing new Hamiltonian $K$, leads to the following H-J equation for 
the generating function $S(x^{\alpha},\gamma_{\beta},\lambda)$:
\be
\frac{1}{2}\, g^{\alpha\beta}\, \frac{\partial S}{\partial x^{\alpha}}
\frac{\partial S}{\partial x^{\beta}} + \frac{\partial S}{\partial \lambda} = 0
\label{HJ_Kerr}
\en
As we have emphasized, the Kerr metric allows full separability of this equation in BL coordinates.
We write
\be
S(x^\alpha,\gamma_\beta,\lambda) = \frac{\mu^2}{2}\lambda -E\,t +L_z\, \phi + S_r(r) + S_\theta (\theta) \equiv
\frac{\mu^2}{2}\lambda + W(x^\alpha,\gamma_\beta)
\label{Sfunction}
\en
The function $W(x^\alpha,\gamma_\beta) $ itself is a generator of a canonical transformation with $K = -\mu^2/2 $.
In any case we find,
\be
 S_r(r) = \int dr\,\frac{\sqrt{R}}{\Delta} \quad \mbox{and} 
\quad S_\theta(\theta) = \int d\theta\, \sqrt{\Theta} 
\en
where $\Delta = r^2 - 2Mr + a^2$.
The potentials $R(r,\alpha_k)$ and $\Theta(\theta,\alpha_k)$, where 
$\alpha_k \equiv \{ -\mu^2/2, E, L_z, Q \} $ are given by the well-known expressions 
(see \cite{MTW},\cite{bardeen}):
\bea
R &=& \left [ (r^2+a^2)\,E -a\,L_z  \right ]^2 -\Delta\,\left [ \mu^2\, r^2 + (L_z -a\,E)^2
+ Q \right ]
\nonumber \\
\nonumber \\
\Theta &=& Q -\cos^2\theta\, \left [ a^2\, (\mu^2 -E^2) + \frac{L_z^2}{\sin^2\theta} 
\right ] .
\label{pot_Kerr}
\ena
The `third' constant $Q$ is the familiar Carter constant and is related to the
original $(r,\theta)$ separation constant ${\cal C}$ as $ Q = {\cal C}^2 -L_z^2 -a^2\,E^2 $.  

Equations of geodesic motion are derived from, 
\bea
 p^\alpha = g^{\alpha\beta}\,\frac{\partial S}{\partial x^\beta} \quad \Rightarrow
\left \{\begin{array}{llll}
& p^t &=& -g^{tt}\, E + g^{t\phi}\, L_z = (r^2+a^2)\,\Delta^{-1}\, T -a\,( a\,E\, \sin^2\theta
-L_z)
\nonumber \\ \nonumber \\
& p^r &=& \pm g^{rr}\,\Delta^{-1}\,\sqrt{R}  = \pm \Sigma^{-1}\,\sqrt{R}
\nonumber \\ \nonumber \\
& p^{\theta} &=& \pm\, g^{\theta\theta}\, \sqrt{\Theta} = \pm \Sigma^{-1}\,\sqrt{\Theta}
\nonumber \\ \nonumber \\
& p^{\phi} &=& g^{\phi\phi}\, L_z -E\, g^{t\phi} = a\,\Delta^{-1}\,T -a\,E 
+ L_z^2/\sin^2\theta
\end{array} \right.\\
\label{EoM_Kerr}
\ena
where $ T = (r^2+a^2)\, E -a\,L_z $ and $ \Sigma = r^2 + a^2\,\cos^2\theta $. Similarly, the covariant momentum 
components are: $ p_t = -E,~ p_r = \pm \sqrt{R}/\Delta,~p_{\theta} = \pm \sqrt{\Theta},~p_{\phi} = L_z $.
Incidentally, we point out that an equivalent `integrated' form of the equations of motion can be 
written by using $ Q^a = \partial S/\partial \gamma_a $, where we can always make the identification 
$\gamma_k = \alpha_k$. 

The periodic character of the motion is revealed when we introduce action-angle canonical variables. 
The actions are defined as (hereafter $i \in \{r,\theta,\phi \} $),
\be
J_i = \oint \, p_i\, dq^i \Rightarrow \left \{\begin{array}{lll}
      & J_r = 2\,\int_{r_p}^{r_a}\, dr\, \sqrt{R}/\Delta  \nonumber \\ \nonumber \\
      & J_{\theta} = 2\, \int_{\theta_n}^{\theta_s}\, d\theta\, \sqrt{\Theta} \nonumber \\
        \nonumber \\
      & J_{\phi} = 2\,\pi\,L_z 
\end{array} \right.
\label{action_Kerr}
\en
and are constant functions $J_i(\alpha_{\kappa})$. In the expression for $J_{\theta}$ we have used $\theta_n, \theta_s$ for
the turning points of the $\theta$-motion. 

The next step is to generate the canonical transformation $ \{x^\alpha,p_\beta \} \Rightarrow \{ w^{\alpha}, I_{\beta}\} $ with 
$I_{\beta} = \{ p_t,J_r,J_\theta,J_\phi \} $. Clearly, $I_{\beta}(\alpha_\kappa) $, or $\alpha_{\kappa}(I_\beta) $.
Since the new momenta $I_\beta $ are constants with respect to $\lambda $ the new Hamiltonian must be $K=K(I_\beta)=K(\alpha_k) $. 
Hence we can set $K=-\mu^2/2 $ and use $W(x^a,I_\beta) $ of (\ref{Sfunction}) as the generating function.

For the new coordinates conjugate to $I_\beta$ we have,
\be
\frac{d w^a}{d\lambda} = \frac{\partial K}{\partial I_\alpha} = \frac{\partial H_\circ}{\partial I_\alpha} \equiv \nu_\alpha \quad
\Rightarrow\quad w^a = \nu_a \lambda + \beta^a
\en
with $\nu_a, \beta^a $ constants. As expected, $w^a $ are angle variables and $\nu_\alpha $ {\it fundamental frequencies}.
These frequencies are related to the orbital frequencies (in BL time) as,
\be
M\Omega_i = \nu_i/\nu_t
\en
The explicit expressions can be found in Ref.~\cite{wolfram}.  
We also have,
\be
w^{\alpha} = \frac{\partial W}{\partial I_{\alpha}} = \frac{\partial W}{\partial \alpha_{\kappa}}\, 
\frac{\partial \alpha_{\kappa}}{\partial I_{\alpha}} 
\label{angle_Kerr}
\en
Expanding eqns.~(\ref{angle_Kerr}), 
\bea
w^t &=&  \nu_t\, \lambda + \beta^t = -2\,\nu_t\,\left [ \frac{\partial S_r}{\partial \mu^2}  +
\frac{\partial S_\theta}{\partial \mu^2} \right ] + t -\left [ \frac{\partial S_r}{\partial E}  +
\frac{\partial S_\theta}{\partial E}   \right ]
\nonumber \\
\nonumber \\
w^r &=& \nu_r\, \lambda + \beta^r = -2\,\nu_r\,\left [ \frac{\partial S_r}{\partial \mu^2}  +
\frac{\partial S_\theta}{\partial \mu^2} \right ]  +  \frac{\partial Q}{\partial J_r} 
\left [ \frac{\partial S_r}{\partial Q}  + \frac{\partial S_\theta}{\partial Q}   \right ]
 \nonumber \\
\nonumber \\
w^\theta &=& \nu_\theta\, \lambda + \beta^\theta = -2\,\nu_\theta\,\left [ \frac{\partial S_r}{\partial \mu^2}  +
\frac{\partial S_\theta}{\partial \mu^2} \right ]  +  \frac{\partial Q}{\partial J_\theta} 
\left [ \frac{\partial S_r}{\partial Q}  + \frac{\partial S_\theta}{\partial Q}   \right ]
\nonumber \\
\nonumber \\
w^\phi &=& \nu_\phi\, \lambda + \beta^\phi = -2\,\nu_\phi\,\left [ \frac{\partial S_r}{\partial \mu^2}  +
\frac{\partial S_\theta}{\partial \mu^2} \right ]  +  \frac{\partial Q}{\partial J_\phi} 
\left [ \frac{\partial S_r}{\partial Q}  + \frac{\partial S_\theta}{\partial Q}   \right ]
+ \frac{1}{2\,\pi}\, \left [ \phi + \frac{\partial S_r}{\partial L_z} + 
\frac{\partial S_\theta}{\partial L_z}   \right ]
\label{angle_Kerr2}
\ena
These are the `integrated' form of the equations of motion and they also provide the $x^{a}(w^k,I_\beta)$ relations
in an implicit form. Eqns.~(\ref{angle_Kerr2}) are fully equivalent to eqns.~(\ref{EoM_Kerr}); this can be easily demonstrated 
by differentiating (say) the $w^r$ and $w^\theta$ equations,
\bea
r^2\,R^{-1/2}\, \dot{r} + a^2\cos^2\,\Theta^{-1/2}\,\dot{\theta} + \frac{1}{2}\, \kappa_r\,
\left [\Theta^{-1/2}\,\dot{\theta} -R^{-1/2}\,\dot{r}  \right ] &=& 1
\nonumber \\
\nonumber \\
r^2\, R^{-1/2}\, \dot{r} + a^2\cos^2\theta\,\Theta^{-1/2}\, \dot{\theta} + \frac{1}{2}\, 
\kappa_\theta\,\left [ \Theta^{-1/2}\,\dot{\theta} -R^{-1/2}\,\dot{\theta} \right ] &=& 1
\label{diff_w}
\ena 
where here an overdot stands for $d/d\lambda$ and
\be
\kappa_i \equiv \frac{1}{\nu_i}\, \frac{\partial Q}{\partial J_i}  
\en  
Combining eqns.~(\ref{diff_w}),
\be
(\kappa_r -\kappa_\theta)\, \left [ \Theta^{-1/2}\,\dot{\theta} - R^{-1/2}\, \dot{r} \right ] = 0
\quad \Rightarrow \quad \Theta^{-1/2}\,\dot{\theta} = R^{-1/2}\,\dot{r} = \sigma(r,\theta)
\en
since $\kappa_r \neq \kappa_\theta $ for non-equatorial orbits. The remaining function $\sigma(r,\theta)$ is 
subsequently specified by substitution into either of eqns.~(\ref{diff_w}). We find 
$ \sigma = (r^2 + a^2\,\cos^2\theta )^{-1} =  \Sigma^{-1} $. Hence, we have arrived at the expected $(r,\theta)$ 
equations of motion. The remaining $(t,\phi)$ equations can be obtained in a similar fashion from the $(w^t,w^\phi)$
equations.

%%%%%%%%%%%%%%%%%%%%%%%%%%%%%%%%%%%%%%%%%%%%%%%%%%%%%%%%%%%%%%%%%%%%

\end{document}